\documentclass[12pt]{article}
\usepackage[utf8]{inputenc}
\usepackage{a4wide}
\usepackage{hyperref}
\usepackage{epsfig}
\usepackage{amsmath}
\usepackage{multirow}
\usepackage{float}
\usepackage{cite}
\usepackage{epsfig}
\usepackage{xspace}
\usepackage{graphics}
\usepackage{relsize}
\usepackage{amsfonts}
\usepackage{ifthen}
\usepackage[normalem]{ulem}	

\def\hide#1{}

\newcommand{\dglap}{{\smaller DGLAP}\xspace}

\newcommand{\ckkwl}{\textnormal{\smaller CKKWL}\xspace} 
\newcommand{\ckkwll}{\textnormal{\smaller CKKW(-L)}\xspace}
\newcommand{\ckkw}{\textnormal{\smaller CKKW}\xspace}
\newcommand{\fxfx}{{\smaller FxFx}\xspace}
\newcommand{\mcnlo}{\textnormal{\smaller MC@NLO\xspace}}
\newcommand{\mlm}{{\smaller MLM}\xspace}

\newcommand{\unlops}{{\smaller UNLOPS}\xspace}

\newcommand{\umeps}{{\smaller UMEPS}\xspace}

\newcommand{\nlthree}{{\smaller NL3}\xspace}
\newcommand{\meps}{{\smaller MEPS}\xspace}
\newcommand{\menlops}{{\smaller MENLOPS}\xspace}
\newcommand{\mepsnlo}{{\smaller MEPS@NLO}\xspace}

\newcommand{\muf}{\mu_F}
\newcommand{\mur}{\mu_R}

\newcommand{\wckkwl}[1]{w_{#1}}
\newcommand{\wumeps}[1]{w_{#1}^\prime}

\newcommand{\ordms}{\ensuremath{\rho_{\mathop{\mathsmaller{\mathsmaller{\mathrm{MS}}}}}}}

\newcommand{\Bornev}[1]{\textnormal{B}_{#1}}
\newcommand{\Bbarev}[1]{\overline{\textnormal{B}}_{#1}}

\newcommand{\Iev}[2]{\int\displaylimits_{\ordms}\widehat{\textnormal{B}}_{#1\rightarrow #2}}
\newcommand{\Tev}[1]{\widehat{\textnormal{B}}_{#1}}

\newcommand{\termX}[2]{\left[~\vphantom{\sum}#1\,\right]_{#2}}

\newcommand{\herwigpp}{{\smaller HERWIG++}\xspace}

\newcommand{\powheg}{{\smaller POWHEG}\xspace}

\newcommand{\sherpa}{{\smaller SHERPA}\xspace}

\newcommand{\pythia}{{\smaller PYTHIA}\xspace}

\newcommand{\pytppp}{{\smaller PYTHIA~8}\xspace}

\newcommand{\as}{\ensuremath{\alpha_{\mathrm{s}}}}

\newcommand{\ord}{\ensuremath{\rho}}
\newcommand{\aux}{\ensuremath{{\boldmath z}}}
\newcommand{\state}[1]{\ensuremath{S_{+#1}}}
\newcommand{\splitP}{\ensuremath{P}}

\newcommand{\done}[1]{}

\def\mrm#1{\mathrm{#1}}

\def\sud#1{\ensuremath{\Delta_{\state{#1}}}}
\def\Pnoem{\ensuremath{\Pi}}
\def\noem#1{\ensuremath{\Pnoem_{\state{#1}}}}

\def\f2d3{\ensuremath{F_2^{\mrm{D}3}}}

\providecommand{\eqref}[1]{eq.~(\ref{#1})\xspace}
\renewcommand{\eqref}[1]{eq.~(\ref{#1})\xspace}

\newcounter{aenumct}

\newcounter{ienumct}

{\end{list}}

\newcounter{enumct}

\def\showcommentsflag{0}

\newcounter{commentcounter}%
\newcommand{\comment}[1]{\ifnum\showcommentsflag > 0%
\addtocounter{commentcounter}{1}%
\Red{\ensuremath{\ddagger^{\arabic{commentcounter}}}}%
\marginpar{\raggedright\tiny\it\Red{\ensuremath{\ddagger^{\arabic{commentcounter}}} #1}}
\fi%
}
\newcommand{\commentdel}[2]{\ifnum\showcommentsflag > 0%
\Red{\sout{#1}}\comment{#2}%
\fi
}
\newcommand{\commentadd}[2]{\ifnum\showcommentsflag > 0%
\comment{#2}\Red{#1}%
\else
#1
\fi
}
\newcommand{\commentchange}[3]{\ifnum\showcommentsflag > 0%
\Red{\sout{#2}}\comment{#3}\Red{#1}%
\else
#1
\fi
}
\newcommand{\nocomment}[1]{\ifnum\showcommentsflag > 0%
{\tiny\it\Red{\{#1}\}}
\fi%
}
\newcommand{\nocommentdel}[1]{\ifnum\showcommentsflag > 0%
\Red{\sout{#1}}%
\fi
}
\newcommand{\nocommentadd}[1]{\ifnum\showcommentsflag > 0%
\Red{#1}%
\else
#1
\fi
}
\newcommand{\nocommentchange}[2]{\ifnum\showcommentsflag > 0%
\Red{\sout{#2}}\Red{#1}%
\else
#1
\fi
}

\begin{document}

\begin{titlepage}

\makeatletter
\def\@fnsymbol#1{\ensuremath{\ifcase#1\or *\or 
   \mathsection\or \mathparagraph\or \|\or **\or \dagger\dagger
   \or \ddagger\ddagger \else\@ctrerr\fi}}
\makeatother

\begin{flushright}
LU TP 15-10\\
March 2015
\end{flushright}
\vfill
\begin{center}
{\Large\bf Tutorial Note on Merging Matrix Elements\\with Parton Showers}
\vfill
{\bf Thomas Rössler}\\[0.3cm]
{Department of Astronomy and Theoretical Physics, Lund University,\\
S\"olvegatan 14A, SE 223-62 Lund, Sweden}
\end{center}
\vfill
\begin{abstract}
In this short note, I introduce to essential conceptual features and main building blocks of matrix element merging algorithms, operating on fixed order calculations both at leading order and next-to-leading order. The intention is purely pedagogical, i.e. to familiarize the reader with the essential basic concepts in a concise way, thus serving as an introduction to beginners and other interested readers. Unitarization is discussed briefly. The tutorial is highly biased towards transverse momentum ordered parton showers, and in particular towards merging schemes as they are implemented in the \pytppp general purpose Monte Carlo generator.
\end{abstract}

\vfill
\vfill

\end{titlepage}

\sloppy



\newpage
\tableofcontents
\bigskip \bigskip \bigskip

\section{Hors d'œuvre}
The present note shall provide a pedagogical and concise introduction to matrix element merging algorithms. It is intended to focus on the main ideas and building blocks. The schemes discussed are dubbed \ckkwll, \umeps, \nlthree and \unlops (cited further below each in their respective section). With the exception of \nlthree, these algorithms are implemented in the \pytppp \cite{Sjostrand:2007gs,1410.3012} event generator and are ready to be used. It is thus also natural that the discussion of merging here comes with a certain bias. For a wider overview over different implementations, taking into account also those available in other event generators, such as e. g. \meps \cite{hep-ph/0205283}, \menlops \cite{1004.1764} and \mepsnlo \cite{1207.5030} in \sherpa \cite{0811.4622}, as well as for phenomenological implications, I strongly recommend a simultaneous study of other literature. 

For practical usage of the merging algorithms, e. g. for phenomenological studies, a consultation of the \pytppp introduction \cite{1410.3012}  and specifically the online manual \cite{PythiaOnline} can be strongly recommended.\footnote{Note that also the \mlm merging scheme \cite{hep-ph/0611129} is available in \pythia. The discussion of that scheme has been omitted to avoid confusion, in particular related to a different "order of steps" as compared to the algorithms discussed here. Similarly, \fxfx \cite{1209.6215} is available as a generalization to NLO.}

The different merging algorithms are discussed on a conceptual level, the text focuses on what the different schemes are able to provide as well as the essential key concepts behind the operations. For explicit structure of the full algorithms, i. e. including the handling of data samples, the reader may be referred to the original publications (cited further below each in their respective section). After discussing the algorithms, I close the note with a short discussion of the terminology \emph{merging} and \emph{matching}.

The note does not contribute anything conceptually new. Its benefit lies fully in its pedagogical structure and clarity, together with its conciseness. I hope it will find its way to readers who will thereby become motivated to both employ as well as further improve QCD corrections in the field of high-energy particle physics.

\section{Motivation}
The merging of matrix elements with parton showers is the art of combining matrix element calculations with different jet multiplicities into a consistent combined framework and to consistently interface them with parton shower resummation. Parton showers provide a longstanding method to do all-order resummation inside of a Monte-Carlo calculation and can model the structure of the soft-collinear divergence of QCD. The parton shower generates consecutive splittings of final state partons by means of expressions which are generally called splitting functions, thus generating a cascade of an increasing number of final state partons. One of the most common choices for splitting functions are the \dglap splitting kernels which are based on collinear factorization of matrix elements. Thus the parton shower alone can, even when being interfaced with only one single Born-type matrix element, give rise to higher jet multiplicities in the event data by generating emissions hard enough to pass the jet definition cut. In this way it properly accounts for the inclusive nature of the matrix element and resums higher-order terms in the strong coupling expansion of the perturbative series.
Fixed-order calculations based on hard matrix elements on the other hand provide a more accurate description for a certain fixed final state parton multiplicity, in particular for hard emissions, i.e. in the phase space regions of well-separated final state partons.
Matrix element merging methods make it possible to combine multiple fixed-order calculations in a consistent picture with the parton shower. The methods are designed to properly account for potential double-counting of final state configurations and to provide Sudakov form factors assigning the necessary no-emission probabilities to make matrix elements exclusive in the desired way. All merging schemes have in common to perform a division of the emission phase space into mainly two distinct regions: One region where the parton shower is used and another region for the hard matrix elements. We refer to the cut between these two regions as the \emph{merging scale}.

\section{Merging scale definition}

For performing a consistent merged calculation, one needs to specify a certain merging scale definition, i.e. what kind of quantity to use as a merging scale cut. For usage with \pytppp, is it a natural choice and advantageous to use a quantity directly related to the evolution variable of the \pytppp parton shower.
The \pytppp shower uses \dglap functions, but kinematically produces $2 \rightarrow 3$ splittings where the third particle is involved as so-called \emph{recoiler} (also often referred to as \emph{spectator}) to ensure energy-momentum conservation. Since the \pytppp evolution scale is a relative $p_T$ distance of the emission, dependent on momenta of three particles, its concept can easily be generalized to three arbitrary partons of any final state and thus also be used on event data from a matrix element calculation.
Already at this stage, I would like to anticipate one fundamental question that all of the merging algorithms will later have to address:

"Given a certain matrix element, how would the parton shower have generated this final state?"

The actual importance of this question will be illustrated more thoroughly further below, but nota bene that this generalization of parts of the parton shower machinery to a matrix element calculation is apparent already here. Given an ordered set of three particles $i,j,k$ out of a matrix element calculation, the corresponding \pythia evolution variable can be calculated. Minimizing this quantity over all ordered triplets of the final state will yield the merging scale definition $\ord_{MS}$ which will be used in the remainder. Thus a parton level final state is said to be above the merging scale if 

\begin{align}
\label{eq:min-pythia-pt-def}
&    t ~=~ \min
        \left[ \ord_{\{i,j,k\}} \right]
    \qquad
    \textnormal{where $i \in$ \{final state partons\},}\\
& \qquad\qquad\qquad\qquad\qquad\!
    \textnormal{and $j \in$ \{final and initial state partons\},}\nonumber\\
& \qquad\qquad\qquad\qquad\qquad\!
    \textnormal{and $k \in$ \{final and initial state partons\}}\nonumber
\end{align}
with the parton separation function
\begin{eqnarray}
\label{eq:pythia-pt-def}
    \ord_{ijk} ~=~
    \begin{cases}
      z_{ijk}(1-z_{ijk})Q_{ij}^2 &
         \textnormal{if the radiator $j$ is a final state parton, and}\\
      &  Q_{ij}^2 = (p_i + p_j)^2 
         ~, \quad
         z_{ijk} = \frac{x_{i,jk}}{x_{i,jk}+x_{j,ik}} \\
      &  x_{i,jk} = \frac{2 p_i(p_i+p_j+p_k)}{(p_i+p_j+p_k)^2}\\
      (1-z_{ijk})Q_{ij}^2 &
         \textnormal{if the radiator $j$ is an initial state parton, and}\\
      &  Q_{ij}^2 = -(p_i - p_j)^2 
         ~, \quad
         z_{ijk} = \frac{(p_i - p_j + p_k)^2}{(p_i+p_k)^2}\\
    \end{cases}
\end{eqnarray}
is above the cut $\ord_{MS}$, i.e. all jets are well separated and considered to be resolved jets, and will correspondingly said to be below the merging scale if $t < \ord_{MS}$, i.e. at least two partons are in the phase space region to be populated by the parton shower.
Note also that this merging scale definition can be used as a cut to regularize the fixed-order input.

\section{\ckkwll merging}
This section is meant to provide a schematic introduction to \ckkwll \cite{Catani:2001cc,Lonnblad:2011xx}. The \mbox{\ckkwll} merging schee combines several leading-order calculations and interfaces them with the shower. First, I will give a short overview over essential features of the \ckkwl algorithm. Further below, some remarks will be made about the differences between \ckkw and Lönnblad's implementation.

Since ultimatively the evaluation of observables is the main interest in any high energy physics Monte Carlo, we start by considering the \ckkwl-based prediction for a general observable given by
\begin{eqnarray}
\langle \mathcal{O} \rangle
 &=& \int d\phi_0\Bigg\{ \mathcal{O}({\state{0j}}) \Bornev{0}\wckkwl{0}
 + \int \mathcal{O}({\state{1j}}) \Bornev{1}\wckkwl{1}
 + \int\!\!\!\int \mathcal{O}({\state{2j}})\Bornev{2}\wckkwl{2} + \\
 \nonumber
&& \qquad\qquad
 \dots
 ~+~ \int\!\dots\!\int \mathcal{O}({\state{Nj}})\Bornev{N}\wckkwl{N}~
\Bigg\} \nonumber\\
&=&
  \sum_{n=0}^N \int d\phi_0 
  \int\!\dots\!\int
  \mathcal{O}({\state{nj}})~
  \Bornev{n}\wckkwl{n}\label{eq:ckkwl}
~,
\end{eqnarray}
where $\Bornev{n}$ denotes the differential leading order cross-section for the process of interest with $n$ additional partons. These cross-sections are multiplied by weights given by
\begin{eqnarray}
\wckkwl{n} &=& 
 \tfrac{x_0^+f_0^+(x_0^+,\ord_0)}{x_n^+f_n^+(x_n^+,\muf)}
 \tfrac{x_0^-f_0^-(x_0^-,\ord_0)}{x_n^-f_n^-(x_n^-,\muf)}
 \times \left(\prod_{i=1}^{n}
  \tfrac{x_i^+f_{i}^+(x_i^+,\ord_i)}{x_{i-1}^+f_{i-1}^+(x_{i-1}^+,\ord_{i-1})}
 ~\tfrac{x_i^-f_{i}^-(x_i^-,\ord_i)}{x_{i-1}^-f_{i-1}^-(x_{i-1}^-,\ord_{i-1})}
  \right)\nonumber\\
&&\times\left(\prod_{i=1}^{n}\tfrac{\as(\ord_i)}{\as(\mur)}\right)
  \times\left(\prod_{i=1}^{n}\noem{i-1}(x_{i-1},\ord_{i-1},\ord_i)\right)
  \times\noem{n}(x_n,\ord_n,\ordms)\label{eq:ckkwl-wgt-old}\\
&=&  \tfrac{x_{n}^+f_{n}^+(x_{n}^+,\ord_n)}{x_{n}^+f_{n}^+(x_{n}^+,\muf)}
     \tfrac{x_{n}^-f_{n}^-(x_{n}^-,\ord_n)}{x_{n}^-f_{n}^-(x_{n}^-,\muf)}\nonumber\\
&&\times 
   \prod_{i=1}^{n} \Bigg[\frac{\as(\ord_i)}{\as(\mur)}
     \tfrac{x_{i-1}^+f_{i-1}^+(x_{i-1}^+,\ord_{i-1})}
          {x_{i-1}^+f_{i-1}^+(x_{i-1}^+,\ord_i)}
     \tfrac{x_{i-1}^-f_{i-1}^-(x_{i-1}^-,\ord_{i-1})}
          {x_{i-1}^-f_{i-1}^-(x_{i-1}^-,\ord_i)}
\noem{i-1}(x_{i-1},\ord_{i-1},\ord_i)\Bigg]\nonumber\\
&&\times\,\,\noem{n}(x_n,\ord_n,\ordms)
  ~,\label{eq:ckkwl-wgt}
\end{eqnarray}
with the exception that the last Sudakov form factor $\noem{n}(x_n,\ord_n,\ordms)$ is not present for the highest jet multiplicity $N$.

The $\ord_i$ are reconstructed emission scales of successive emissions in the definition of the shower evolution variable. As has been intimated above, the merging algorithm processes matrix elements by trying to answer the question: ``How would the parton shower have generated this final state?''. It will try to construct a sensible parton shower history, ideally starting from the 0-jet Born matrix element followed by $n$ successive emissions with strictly ordered emission scales, thus being a valid event that could have equally well been generated by the transverse momentum ordered \pytppp shower in precisely that order.

The \ckkwl procedure then reweighs the matrix elements with parton shower weights that contain PDF ratios, the running of the strong coupling used in the parton shower as well as Sudakov form factors. The latter play a crucial role in making the events exclusive with respect to the merging scale.

In an ordinary parton shower, the Sudakov form factors naturally emerge by the nature of the veto algorithm. This feature can also be exploited here, by running a trial shower on the reclustered event. Then, respective form factors can be generated by the following ``modified'' veto algorithm for each emission step:

Assuming an emission at a reconstructed scale $\ord_i$, we need to generate a Sudakov of type $\noem{i-1}(x_{i-1},\ord_{i-1},\ord_i)$. Starting from the reclustered state, perform a trial emission. Now distinguish between the two cases

I. The trial emission has been generated at a higher scale than the reconstructed one. In that case, veto the event.

II. The trial emission has been generated at a lower scale than the reconstructed one. In this case, keep the actual emission of the parton shower history and continue processing the event.

Nota bene that the actual scale of the trial emission is discarded in any case.

The ``last'' Sudakov factor $\noem{n}(x_n,\ord_n,\ordms)$ is generated using the same strategy by comparison to $\ord_{MS}$. As already mentioned, this factor is not generated in the highest jet multiplicity due to the benefit from keeping this type of events inclusive. Since there are no corresponding matrix elements for $N+1$ partons above the merging scale, these events are not double-counted and thus should not be removed in order to avoid the creation of a hole in the emission phase space.

Let me mention only a few of the details and complications regarding the algorithm and the construction of a parton shower history and outline the solutions that were chosen.

- Disconnected diagrams. Taking a final state of a matrix element and clustering its particles consistent with quantum numbers does not necessarily lead to a connected Feynman diagram, thus not yielding at all a valid emission history for the process. These disconnected diagrams are discarded, since connected emission histories can always be created.

- Incomplete histories: The state cannot be completely clustered to the desired parton multiplicity and the history is incomplete. The clustering algorithm thus just leads to an unreasonable underlying Born. Incomplete paths are not considered if a (ordered or unordered) history can be generated.

- Unordered emissions: Histories can be generated with scales that are not $p_T$ ordered in the order of consective splittings, thus could not have been generated by the $p_T$ ordered \pytppp parton shower in this way. Unordered histories are avoided as long as ordered ones can be created. Otherwise, an approximative treatment of Sudakov no-emission probabilities is performed.
More details can be found in \cite{Lonnblad:2011xx}.

Due to the nature of the algorithm, the total cross-section in this scheme acquires a dependence on the merging scale $\ord_{MS}$. This is clear, because weights are added to the calculation in the matrix element region being different from the weights that would have been obtained showering the Born matrix element only.\footnote{These effects are sometimes denoted "unitarity violations" in the literature since they violate the parton shower unitarity.}

The merging scheme was invented independently by \ckkw and Lönnblad. A thorough comparison of both publications still evinces certain differences in the procedure of which at least the following two should be mentioned:

$\rightarrow$ \ckkw apply analytic Sudakov factors to their matrix elements whereas Lönnblad utilizes a trial shower to generate this factor by means of the veto algorithm

$\rightarrow$ \ckkw perform their clustering by using the longitudinally invariant kT-algorithm. Their procedure is thus fully deterministic whereas Lönnblad generates all possible parton shower histories, followed by a probabilistic selection.

\section{The concept of unitarization}
This section adresses the issue of a unitarized merging. Starting from the \ckkwl algorithm, the necessary steps will be taken to extend the algorithm to a variant that conserves the parton shower unitarity. Low multiplicity states are supplemented with approximate higher orders from higher multiplicity states in order to retain the $n$-jet leading-order inclusive cross-section. The method has been dubbed \umeps \cite{Lonnblad:2012ng} by the authors of the respective \pytppp implementation.\footnote{A unitarized merging algorithm has been suggested independently in \cite{Platzer:2012bs} for \herwigpp.}
The \umeps event weight is given by
\begin{eqnarray}
\wumeps{n}
&=&  \tfrac{x_{n}^+f_{n}^+(x_{n}^+,\ord_n)}{x_{n}^+f_{n}^+(x_{n}^+,\muf)}
     \tfrac{x_{n}^-f_{n}^-(x_{n}^-,\ord_n)}{x_{n}^-f_{n}^-(x_{n}^-,\muf)}
   \nonumber\\
&&\times
     \prod_{i=1}^{n} \Bigg[\tfrac{\as(\ord_i)}{\as(\mur)}
     \tfrac{x_{i-1}^+f_{i-1}^+(x_{i-1}^+,\ord_{i-1})}
          {x_{i-1}^+f_{i-1}^+(x_{i-1}^+,\ord_i)}
     \tfrac{x_{i-1}^-f_{i-1}^-(x_{i-1}^-,\ord_{i-1})}
          {x_{i-1}^-f_{i-1}^-(x_{i-1}^-,\ord_i)}
  \noem{i-1}(x_{i-1},\ord_{i-1},\ord_i)\Bigg]
  ~.\label{eq:umeps-wgt}
\end{eqnarray}

This differs from the CKKWL weight (equation \ref{eq:ckkwl-wgt}), as it does not contain the ``last'' Sudakov factor $\noem{n}(x_n,\ord_n,\ordms)$.\footnote{Thus, note also $\wckkwl{N}=\wumeps{N}$ where $N$ denotes the highest included multiplicity.} In \umeps, this piece is instead generated by the higher-order multiplicities. The principle is motivated by the mechanism how an ordinary parton shower retains the inclusive cross-section. \umeps makes this cancellation explicit: The emission part of the Sudakov form factor is generated mainly from the next-higher multiplicity after the last emission has been integrated out. Since this part comes with a minus sign, the positive-weight contribution coming from a jet that has been emitted at a scale $\ord_{em}$ is cancelled with a negative contribution where no emission between a maximal scale $\ord_{max}$ and $\ord_{em}$ has taken place. For example, events with two resolved jets can exactly cancel with events having one unresolved jet in the one-jet state.
Note that this subtractive treatment can be conveniently implemented using the parton shower history that has already been used for \ckkwl. In the \umeps scheme, the prediction for an observable reads

\allowdisplaybreaks
\begin{eqnarray}
\label{eq:umeps-observable}
\langle \mathcal{O} \rangle
 &=& \int d\phi_0\Bigg\{ \mathcal{O}({\state{0j}})\left[ \Tev{0}
 ~-~ \Iev{1}{0}
 ~-~ \Iev{2}{0}
 ~-~
\dots
 ~-~ \Iev{N}{0}~\right]\nonumber\\
&&\qquad\quad
 + \int \mathcal{O}({\state{1j}}) \left[
     \Tev{1}
 ~-~ \Iev{2}{1}
 ~-~
\dots
 ~-~ \Iev{N}{1}~\right]\nonumber\\
&&\qquad\quad
 +\phantom{\int} \dots\nonumber\\
&&\qquad\quad
 + \int\!\dots\!\int \mathcal{O}({\state{N-1j}}) \left[
     \Tev{N-1}
 ~-~ \Iev{N}{N-1}~\right]
\nonumber\\
&&\qquad\quad
 + \int\!\cdots\!\int \mathcal{O}({\state{Nj}})~\Tev{N} ~\Bigg\}\nonumber\\
&=&
  \sum_{n=0}^N \int d\phi_0 
    \int\!\dots\!\int
  \mathcal{O}({\state{nj}})~\left\{
  \Tev{n}
 ~-~ 
  \sum_{i=n+1}^{N}
  \Iev{i}{n}
  ~\right\}
~.
\end{eqnarray}
\allowdisplaybreaks[0]
where $\Bornev{n}\wumeps{n} ~=~ \Tev{n}$ has been used as concise notation and
\begin{eqnarray}
\Iev{n}{m}~
 ~=~ \int d^{n-m}\phi~\Bornev{n}\wumeps{n}
\end{eqnarray}
denotes the subtractive terms. As a subtlety, note that further integrations can be necessary since the merging scale measure of an event might drop below $\ordms$ due to the correct undoing of recoil effects, thus the additional terms $\Iev{n}{m}$ with $m \neq n-1$ in equation (\ref{eq:umeps-observable}).

In \umeps, the inclusive $n$-jet cross-section (for $n < N$) is not made exclusive by its own Sudakov resummation coming from $n$-jet events being processed by the shower. Instead it is made exclusive by the subtractive treatment of the $n+1$-jet inclusive cross-section. This feature can also be looked at by considering a recursive expansion of the "last" Sudakov form factor as
\begin{eqnarray}
  \label{eq:sud}
  \sud{n}(\ord_n,\ordms)&=
  &\exp\left[-\int_{\ordms}^{\rho_n}d\rho\int d\aux
    \as(\rho) \splitP_i(\rho,\aux)\right]~\\
  &=&1-\int_{\ordms}^{\rho_n}d\rho\int d\aux
    \as(\rho) \splitP_i(\rho,\aux)\sud{n}(\ord_n,\ord)~.
\end{eqnarray}

The step from \ckkwl to \umeps can then also be understood as the following transition: Take the \ckkwl-weighted expression, expand the Sudakov factor according to equation \ref{eq:sud}. The cross-section will now contain the $n$-jet maxtrix element times a splitting function. Go over to \umeps by promoting this product to an exact $(n+1)$-jet matrix element, thus ``undoing`` the collinear factorization.

Loosely speaking, the philosophy of unitarization is ''we subtract what we add``: Since an integrated version of the terms from the higher multiplicity matrix elements is subtracted again, the algorithm preserves the inclusive $n$-jet leading-order cross-section by construction.

\section{\nlthree}
The \nlthree merging scheme \cite{Lavesson:2008ah} allows to incorporate one or several NLO calculations into the merging framework without inducing double-counting and still retaining resummation effects. It can be considered as an NLO generalization of \ckkwl. The discussion will be kept rather short. It is mainly shown for instructive reasons as an intermediate stage to the \unlops merging scheme which is discussed in section \ref{sec:unlops}.

The prediction for an observable in \nlthree merging is given by

\begin{eqnarray}
\label{eq:nl3-full-inclusive-main-text}
\langle \mathcal{O} \rangle
 &=& \sum_{m=0}^M \int d\phi_0 \int\!\cdots\!\int 
     \mathcal{O}(\state{mj})
     ~\Bigg\{ \termX{\Bornev{m}\wckkwl{m}}{-m,m+1} + \Bbarev{m} 
 -\int_s \Bornev{m+1\rightarrow m} \quad\Bigg\}\nonumber\\
&+&
  \sum_{n=M+1}^N
  \int d\phi_0 \int\!\cdots\!\int 
  \mathcal{O}(\state{nj})
  \Bornev{n} \wckkwl{n}
\end{eqnarray}
where $M < N$ NLO calculations have been added to our $N$ tree-level calculations and $\Bbarev{m}$ denotes an inclusive NLO calculation. The bracket notation $\termX{\Bornev{m}\wckkwl{m}}{m,m+1}$ projects out certain powers of the $\alpha_s$ expansion, i.e. $\alpha_s^m$ and $\alpha_s^{m+1}$, whereas $\termX{\Bornev{m}\wckkwl{m}}{-m,m+1}$ means that corresponding powers of $\alpha_s$ are explicitly removed.
The aim of the algorithm is to provide NLO-correct results for exclusive $n$-jet observables while simultaneously retaining the parton shower resummation for all higher orders. Compared to the \ckkwl prediction (equation \ref{eq:ckkwl}), the important change is thus that $M$ NLO calculations have been added with simultaneous removal of corresponding terms from the tree-level samples which would be double-counted otherwise. Merging schemes naturally act on exclusive cross-sections. Note that the inclusive NLO calculation is made exclusive by the inclusion of a phase space subtraction sample.

\section{\unlops}
\label{sec:unlops}
\unlops \cite{Lonnblad:2012ix} restores the parton shower unitarity for NLO merging in a way similar to \umeps in the leading-order case. It is now the NLO $n$-jet inclusive cross-section which we wish to conserve while simultaneously obtaining NLO-correct results for exclusive $n$-jet observables. The strategy is again to ''subtract what we add`` from the lower multiplicities.

As for \nlthree, we show the prediction formula for inclusive NLO input.

\begin{eqnarray}
\label{eq:unlops-full-inclusive-appendix}
\langle \mathcal{O} \rangle
 &=&
 \sum_{m=0}^{M-1}~ \int d\phi_0 \int\!\cdots\!\int 
     \mathcal{O}(\state{mj})
     ~\Bigg\{
   \Bbarev{m}
 + \termX{\Tev{m}}{-m,m+1}
\nonumber\\
&&\qquad\qquad\qquad\quad~
 - \sum_{i=m+1}^{M} \int_s\Bbarev{i\rightarrow m}
 - \sum_{i=m+1}^{M} \termX{\Iev{i}{m}}{-i,i+1}
 - \sum_{i=M+1}^{N} \Iev{i}{m}
~\Bigg\}\nonumber\\
&+&
   \int d\phi_0 \int\!\cdots\!\int 
   \mathcal{O}(\state{Mj})\Bigg\{~
   \Bbarev{M}
 + \termX{\Tev{M}}{-M,M+1}
 - \sum_{i=M+1}^{N} \Iev{i}{M}~
   \Bigg\}\nonumber\\
&+&
  \sum_{n=M+1}^N
  \int d\phi_0 \int\!\cdots\!\int 
  \mathcal{O}(\state{nj})~\left\{ \Tev{n} - \sum_{i=n+1}^{N} \Iev{i}{n} ~\right\}
\end{eqnarray}

Again, $M < N$ NLO calculations have been added to our $N$ tree-level calculations. Compared to \umeps (equation \ref{eq:umeps-observable}), inclusive NLO calculations $\Bbarev{m}$ have been introduced and, as for \nlthree, respective terms in the strong coupling expansion of the tree-level samples have been removed. Note that, if more than one NLO calculation is used, the restoration of unitarity requires

I. an additional subtractive treatment with integrated emission also for the inclusive $n$-jet NLO calculations ($n \neq 0$)

II. that the terms that have been removed from the $n$-jet tree-level calculation are similarly removed from their subtractive treatments in the lower-multiplicity bins (for all $n$ that are supplemented with an NLO calculation)

Higher multiplicity tree-level calculations will improve the description of exclusive $n$-jet observables while additionally the total inclusive $m$-jet cross-sections remain unchanged from the ones generated by the NLO calculations.
NLO calculations with additional jets, in particular with multiple jets, are still not available for a lot of processes of interest. A common choice therefore is to supplement a single NLO 0-jet calculation with higher orders from tree level samples.\footnote{The reader paying close attention will note that formula \ref{eq:unlops-full-inclusive-appendix} will in this case break down to the \umeps formula \ref{eq:umeps-observable} with the replacement $\Tev{0} \rightarrow \Bbarev{0}$.}

\section{A short remark on the terminology of merging and matching}
In particular in the older literature, a certain confusing inconsistency was present related to the terms \emph{merging} and \emph{matching}. As happens rather frequently in the etymological history of new words, it took time until a clear and well-defined meaning could crystallize out for each of these two terms. In this section, I try to clarify them in the way they are used by most people these days.

Two common standard approaches are frequently used in the implementation of NLO calculations: the \powheg \cite{hep-ph/0409146} method and \mcnlo \cite{hep-ph/0204244}. The two approaches mainly differ in their subtractive treatment which is beyond the scope of this note. Formally, they are equivalent to the first order in the strong coupling expansion, differing only in the higher order terms. The term \emph{matching} refers to the question how a (single) NLO calculation can be consistently interfaced with a parton shower. This is non-trivial, obviously process-dependent, and different for both NLO approaches. \emph{Merging} on the other hand addresses the question how \emph{several} fixed-order calculations can be combined. As is readily apparent from the discussion above, merging can be constructed in a generic process-independent way.

Merging and matching are of course - even with these clear definitions - intimately related, and address essentially the same problems (e. g. avoiding double-counting of terms). Nevertheless, one should have in mind that this terminology has not been that clear all the time, and were thus in the past often (regarded from today's perspective) confused or even used as synonyms for each other.

\newpage

\bibliographystyle{utcaps}
\bibliography{intro-references}

\providecommand{\href}[2]{#2}\begingroup\raggedright\begin{thebibliography}{10}

\bibitem{Sjostrand:2007gs}
T.~Sjöstrand, S.~Mrenna, and P.~Skands,
  \href{http://dx.doi.org/10.1016/j.cpc.2008.01.036}{{\em Comput. Phys.
  Commun.} {\bf 178} (2008)  852--867},
\href{http://arxiv.org/abs/0710.3820}{{\tt arXiv:0710.3820 [hep-ph]}}.

\bibitem{1410.3012}
T.~Sjöstrand, S.~Ask, J.~R. Christiansen, R.~Corke, N.~Desai, {\em et al.},
\href{http://arxiv.org/abs/1410.3012}{{\tt arXiv:1410.3012 [hep-ph]}}.

\bibitem{hep-ph/0205283}
F.~Krauss, \href{http://dx.doi.org/10.1088/1126-6708/2002/08/015}{{\em JHEP}
  {\bf 0208} (2002)  015},
\href{http://arxiv.org/abs/hep-ph/0205283}{{\tt arXiv:hep-ph/0205283
  [hep-ph]}}.

\bibitem{1004.1764}
K.~Hamilton and P.~Nason, \href{http://dx.doi.org/10.1007/JHEP06(2010)039}{{\em
  JHEP} {\bf 1006} (2010)  039},
\href{http://arxiv.org/abs/1004.1764}{{\tt arXiv:1004.1764 [hep-ph]}}.

\bibitem{1207.5030}
S.~Hoeche, F.~Krauss, M.~Schonherr, and F.~Siegert,
  \href{http://dx.doi.org/10.1007/JHEP04(2013)027}{{\em JHEP} {\bf 1304} (2013)
   027},
\href{http://arxiv.org/abs/1207.5030}{{\tt arXiv:1207.5030 [hep-ph]}}.

\bibitem{0811.4622}
T.~Gleisberg, S.~Hoeche, F.~Krauss, M.~Schonherr, S.~Schumann, {\em et al.},
  \href{http://dx.doi.org/10.1088/1126-6708/2009/02/007}{{\em JHEP} {\bf 0902}
  (2009)  007},
\href{http://arxiv.org/abs/0811.4622}{{\tt arXiv:0811.4622 [hep-ph]}}.

\bibitem{PythiaOnline}
\url{http://home.thep.lu.se/~torbjorn/pythia82php/Welcome.php}.

\bibitem{hep-ph/0611129}
M.~L. Mangano, M.~Moretti, F.~Piccinini, and M.~Treccani,
  \href{http://dx.doi.org/10.1088/1126-6708/2007/01/013}{{\em JHEP} {\bf 0701}
  (2007)  013},
\href{http://arxiv.org/abs/hep-ph/0611129}{{\tt arXiv:hep-ph/0611129
  [hep-ph]}}.

\bibitem{1209.6215}
R.~Frederix and S.~Frixione,
  \href{http://dx.doi.org/10.1007/JHEP12(2012)061}{{\em JHEP} {\bf 1212} (2012)
   061},
\href{http://arxiv.org/abs/1209.6215}{{\tt arXiv:1209.6215 [hep-ph]}}.

\bibitem{Catani:2001cc}
S.~Catani, F.~Krauss, R.~Kuhn, and B.~R. Webber, {\em JHEP} {\bf 11} (2001)
  063, \href{http://arxiv.org/abs/hep-ph/0109231}{{\tt hep-ph/0109231}}.

\bibitem{Lonnblad:2011xx}
{L.~L\"onnblad, and S.~Prestel},
  \href{http://dx.doi.org/10.1007/JHEP03(2012)019}{{\em JHEP} {\bf 1203} (2012)
   019},
\href{http://arxiv.org/abs/1109.4829}{{\tt arXiv:1109.4829 [hep-ph]}}.

\bibitem{Lonnblad:2012ng}
L.~L{\"o}nnblad and S.~Prestel,
\href{http://arxiv.org/abs/1211.4827}{{\tt arXiv:1211.4827 [hep-ph]}}.

\bibitem{Platzer:2012bs}
S.~Plätzer, \href{http://dx.doi.org/10.1007/JHEP08(2013)114}{{\em JHEP} {\bf
  1308} (2013)  114},
\href{http://arxiv.org/abs/1211.5467}{{\tt arXiv:1211.5467 [hep-ph]}}.

\bibitem{Lavesson:2008ah}
N.~Lavesson and L.~Lönnblad,
  \href{http://dx.doi.org/10.1088/1126-6708/2008/12/070}{{\em JHEP} {\bf 12}
  (2008)  070},
\href{http://arxiv.org/abs/0811.2912}{{\tt arXiv:0811.2912 [hep-ph]}}.

\bibitem{Lonnblad:2012ix}
L.~Lönnblad and S.~Prestel,
  \href{http://dx.doi.org/10.1007/JHEP03(2013)166}{{\em JHEP} {\bf 1303} (2013)
   166},
\href{http://arxiv.org/abs/1211.7278}{{\tt arXiv:1211.7278 [hep-ph]}}.

\bibitem{hep-ph/0409146}
P.~Nason, \href{http://dx.doi.org/10.1088/1126-6708/2004/11/040}{{\em JHEP}
  {\bf 0411} (2004)  040},
\href{http://arxiv.org/abs/hep-ph/0409146}{{\tt arXiv:hep-ph/0409146
  [hep-ph]}}.

\bibitem{hep-ph/0204244}
S.~Frixione and B.~R. Webber,
  \href{http://dx.doi.org/10.1088/1126-6708/2002/06/029}{{\em JHEP} {\bf 0206}
  (2002)  029},
\href{http://arxiv.org/abs/hep-ph/0204244}{{\tt arXiv:hep-ph/0204244
  [hep-ph]}}.

\end{thebibliography}\endgroup

\end{document}